\documentclass[reprint,superscriptaddress,aps]{revtex4-1}
\usepackage{amsmath}
\usepackage{amssymb}
\usepackage{bm}
\usepackage{braket}
\usepackage{color}
\usepackage[varg]{txfonts}
\usepackage{varwidth}
\usepackage{dcolumn}
\usepackage[breaklinks,colorlinks=true,linkcolor=blue,urlcolor=cyan,citecolor=blue]{hyperref}
\usepackage{graphicx}

\begin{document}

\title{Effect of uniaxial stress on helimagnetic phases in the square-lattice itinerant magnet EuAl$_{4}$}

\author{Masaki Gen}
\email{gen@issp.u-tokyo.ac.jp}
\affiliation{Institute for Solid State Physics, University of Tokyo, Kashiwa 277-8581, Japan}
\affiliation{RIKEN Center for Emergent Matter Science (CEMS), Wako 351-0198, Japan}

\author{Takuya Nomoto}
\affiliation{Department of Physics, Tokyo Metropolitan University, Hachioji, Tokyo 192-0397, Japan}

\author{Hiraku~Saito}
\affiliation{Institute for Solid State Physics, University of Tokyo, Kashiwa 277-8581, Japan}

\author{Taro~Nakajima}
\affiliation{Institute for Solid State Physics, University of Tokyo, Kashiwa 277-8581, Japan}
\affiliation{RIKEN Center for Emergent Matter Science (CEMS), Wako 351-0198, Japan}
\affiliation{Institute of Materials Structure Science, High Energy Accelerator Research Organization, Tsukuba 305-0801, Japan}

\author{Yusuke~Tokunaga}
\affiliation{Department of Advanced Materials Science, The University of Tokyo, Kashiwa 277-8561, Japan}

\author{Rina~Takagi}
\affiliation{Institute for Solid State Physics, University of Tokyo, Kashiwa 277-8581, Japan}

\author{Shinichiro~Seki}
\affiliation{Research Center for Advanced Science and Technology, University of Tokyo, Tokyo 113-8656, Japan}
\affiliation{Department of Applied Physics, The University of Tokyo, Tokyo 113-8656, Japan}

\author{Taka-hisa~Arima}
\affiliation{RIKEN Center for Emergent Matter Science (CEMS), Wako 351-0198, Japan}
\affiliation{Department of Advanced Materials Science, The University of Tokyo, Kashiwa 277-8561, Japan}

\begin{abstract}

We investigate uniaxial-stress effects on the magnetic phase diagram of the square-lattice itinerant magnet EuAl$_{4}$, where strong coupling among spin, lattice, and charge produces a variety of helimagnetic phases, including rhombic and square skyrmion lattices.
Combining resistivity and magnetization measurements with neutron scattering, we find that compressive stresses of only several tens of megapascal along [010] enhance antiferromagnetic character and shorten the magnetic modulation period in the lowest-temperature single-${\mathbf Q}$ spiral state, thereby driving the critical temperatures and fields of multiple phases to higher values.
First-principles calculations show that increasing orthorhombic lattice distortion deforms the Fermi surface relevant to the magnetism, providing compelling evidence that Fermi-surface nesting plays a crucial role in stabilizing the helical magnetic modulations in EuAl$_{4}$.

\end{abstract}

\date{\today}
\maketitle

Uniaxial stress or strain serves as a versatile control knob for symmetry-selective tuning of the coupled lattice, charge, orbital, and spin degrees of freedom in quantum materials.
Unlike hydrostatic pressure, uniaxial stress directly lowers crystal symmetry, thereby reshaping band dispersions and modifying Fermi-surface geometry.
This approach has been utilized to manipulate superconducting transition temperatures \cite{2014_Hic}, to probe multicomponent superconducting states \cite{2025_Gho}, to reveal electronic nematicity \cite{2021_Ike}, to tune the wave vector and onset of charge density wave (CDW) states \cite{2018_Gao, 2018_Kim, 2021_Qia}, and to drive Lifshitz transitions \cite{2019_Sun, 2021_Nic}.
Beyond these itinerant instabilities, it is instructive to recall the long-standing piezoelectric route to electromechanical control, which has seen wide application in sensors and devices such as accelerometers, microphones, and ultrasound receivers \cite{1999_Hae, 2023_Sek}.
Moreover, growing attention has turned to its magnetic analog, piezomagnetism, in which strain couples linearly to magnetic order and can generate a net magnetization \cite{2017_Jai, 2024_Men}.
In noncollinear kagome antiferromagnets, even small lattice distortions modulate the Berry curvature and enable piezomagnetic switching of the anomalous Hall effect \cite{2020_Guo, 2022_Ikh}.

Over the past decade, the effects of uniaxial stress on the stability of magnetic skyrmion lattice (SkL), a topologically nontrivial swirling spin texture, have been extensively explored in chiral magnets such as MnSi \cite{2015_Nii, 2015_Cha, 2017_Fob}, FeGe \cite{2015_Shi, 2020_Bud, 2022_Lit}, and Cu$_{2}$OSeO$_{3}$ \cite{2017_Sek, 2018_Nak}.
In these systems, competition between the Dzyaloshinskii-Moriya (DM) interaction and ferromagnetic (FM) exchange interaction produces helical modulations, and within a narrow window of magnetic field and temperature, a triangular SkL is stabilized due to the entropy effect \cite{2009_Muh, 2011_Yu, 2012_Sek}.
Because SkL stability is exquisitely sensitive to the magnitudes of magnetic anisotropy and the DM interaction, modest uniaxial stress has been shown to dramatically reorganize the SkL phase diagram by tuning these parameters, in both experiments \cite{2015_Nii, 2015_Cha, 2017_Fob, 2015_Shi, 2020_Bud, 2022_Lit, 2017_Sek, 2018_Nak} and theory \cite{2010_But, 2015_Lin, 2017_Kan, 2021_Hog}.
More recently, a SkL phase driven by Ruderman-Kittel-Kasuya-Yosida (RKKY) interaction has been proposed theoretically \cite{2017_Oza, 2017_Hay, 2024_Hay}, and subsequently observed in several centrosymmetric rare-earth based intermetallics such as Gd$_{2}$PdSi$_{3}$ \cite{2019_Kur}, GdRu$_{2}$Si$_{2}$ \cite{2020_Kha}, GdRu$_{2}$Ge$_{2}$ \cite{2024_Yos}, and EuAl$_{4}$ \cite{2022_Tak}.
To date, however, uniaxial-stress effects on the magnetic phase diagrams of these second-generation SkL hosts have not been investigated.

\begin{figure}[t]
\centering
\includegraphics[width=0.92\linewidth]{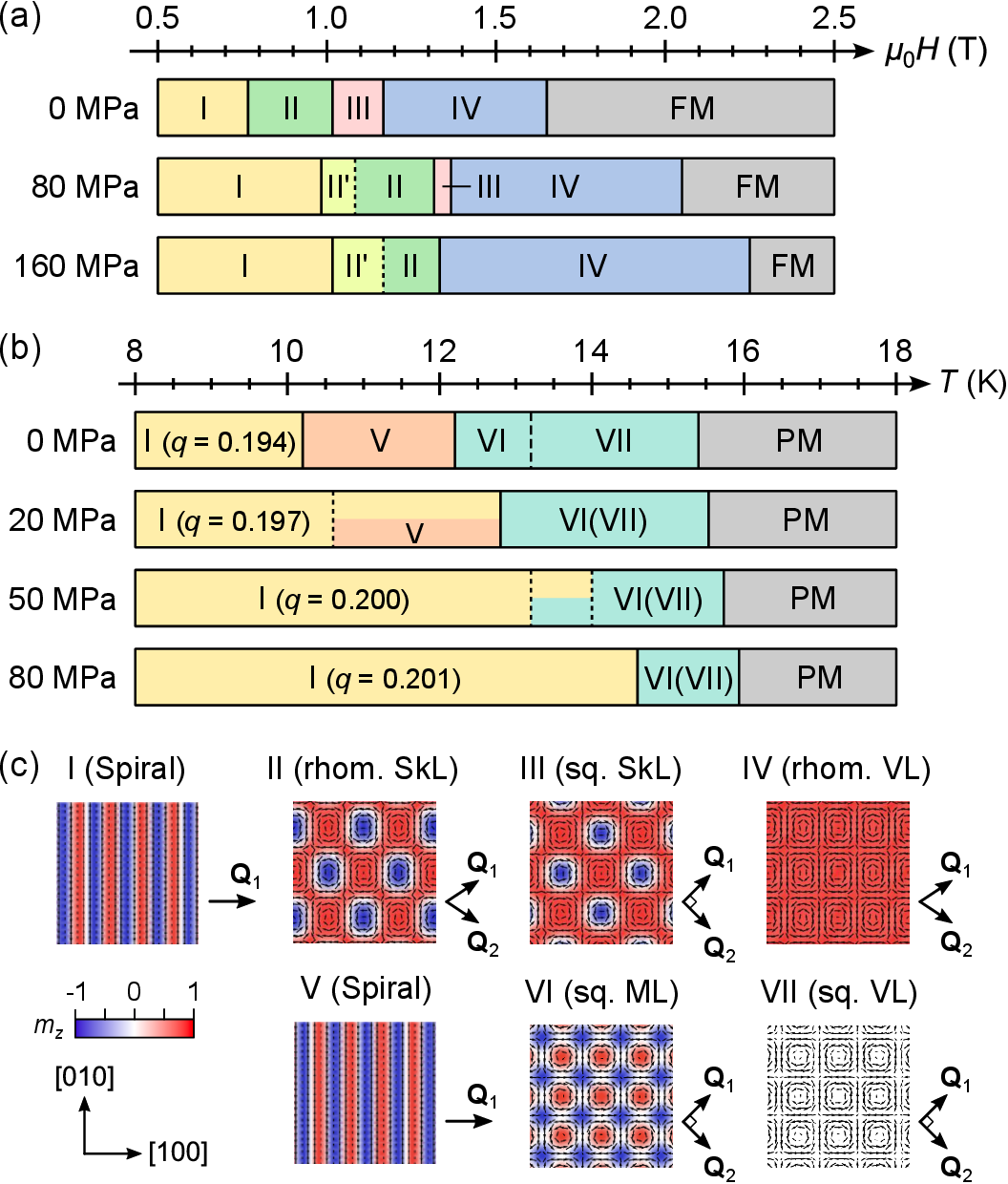}
\caption{(a) Magnetic field--stress phase diagram for $H \parallel [001]$ below 5~K, determined from the present magnetization measurements. (b) Temperature--stress phase diagram at zero magnetic field, determined from the present resistivity measurements and neutron-scattering experiments. The compressive uniaxial stress was applied along the [010] direction. (c) Schematic illustration of the magnetic structures and their modulation vectors in the seven helimagnetic phases for $H \parallel [001]$ in EuAl$_{4}$ \cite{2022_Tak}. The helicities in phases~I and V are opposite to each other \cite{2024_Mia, 2024_Vib}.}
\label{Fig1}
\end{figure}

In this Letter, we report that the magnetic phase diagram of EuAl$_{4}$ can be dramatically tuned by compressive stresses of only several tens of megapascal, as shown in Figs.~\ref{Fig1}(a) and \ref{Fig1}(b).
EuAl$_{4}$ is one of the most complex SkL hosts, exhibiting diverse helimagnetic phases intertwined with a CDW and an orthorhombic lattice distortion \cite{2022_Tak, 2015_Nak, 2019_Shi, 2021_Sha, 2021_Kan, 2022_Mei, 2023_Gen, 2024_Mia, 2024_Vib, 2022_Ram, 2024_Kor, 2025_Kot, 2024_Ni, 2025_Cao, 2025_Suk,  2016_Kob, 2024_Eat}.
At room temperature, the crystal structure is centrosymmetric tetragonal (space group $I4/mmm$), comprising a square lattice of localized Eu$^{2+}$ ions with spin 7/2.
At $T_{\rm CDW} = 145$~K, EuAl$_{4}$ undergoes a CDW transition characterized by a horizontal displacement of Al atoms with an incommensurate modulation vector ${\mathbf Q}_{\rm CDW} \approx (0, 0, 0.17)$ in the reciprocal lattice unit \cite{2015_Nak, 2019_Shi, 2021_Sha, 2021_Kan, 2022_Ram, 2024_Kor, 2025_Kot, 2024_Ni, 2025_Cao, 2025_Suk}.
Upon further cooling below $T_{\rm N1} = 15.4$~K, four magnetic transitions from the paramagnetic (PM) phase to phases VII, VI, V, and I occur within a narrow temperature interval \cite{2022_Tak, 2015_Nak, 2019_Shi, 2021_Sha, 2021_Kan, 2022_Mei, 2023_Gen, 2024_Mia, 2024_Vib}.
The transition at $T_{\rm N3} = 12.2$~K is accompanied by an orthorhombic structural transition with the $B_{1g}$-type distortion \cite{2019_Shi, 2023_Gen}.
With a magnetic field applied along [001], a sequence of transitions from phase I to II, III, and IV takes place before the forced FM state is reached \cite{2022_Tak, 2015_Nak, 2021_Sha, 2022_Mei, 2023_Gen}.
The magnetic structures in phases I and V are single-${\mathbf Q}$ spirals with ${\mathbf Q}_{1}=(0.194, 0, 0)$ and (0.17, 0, 0), respectively, whereas the remaining phases are double-${\mathbf Q}$ states; phases II and III are SkL, phases IV and VII are vortex-antivortex lattices (VL), and phase VI is a meron-antimeron lattice (ML) \cite{2022_Tak, 2021_Kan, 2024_Mia, 2024_Vib} [Fig.~\ref{Fig1}(c)].
Two fundamental modulation vectors in phases III, VI, and VII are ${\mathbf Q}_{1}=(q, q, 0)$ and ${\mathbf Q}_{2}=(q, -q, 0)$ with $q \approx 0.085$, while in phases II and IV, $ {\mathbf Q}_{1}$ and ${\mathbf Q}_{2}$ are slightly tilted toward [100], resulting in a rhombic distortion of the SkL and VL, respectively \cite{2022_Tak}.
Previous dilatometry and resonant x-ray scattering have revealed a strong correlation between the broken four-fold symmetry of the magnetic structures and the orthorhombic lattice distortion \cite{2022_Mei, 2023_Gen}, underscoring the feasibility of controlling the magnetic structures by in-plane uniaxial stress.

\begin{figure}[t]
\centering
\includegraphics[width=0.98\linewidth]{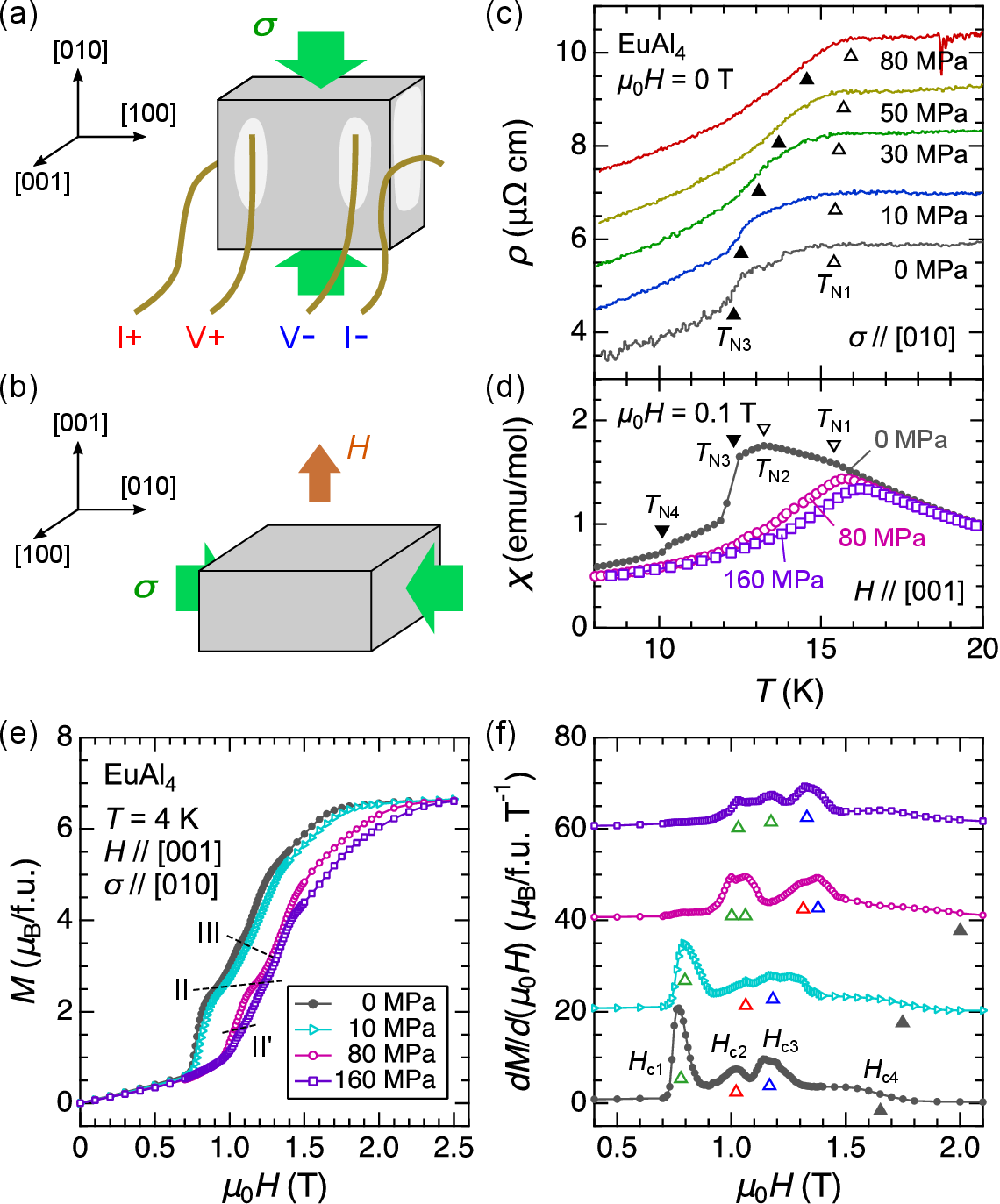}
\caption{(a),(b) Schematic illustrations of the experimental geometry for (a) resistivity and (b) magnetization measurements under compressive stress $\sigma_{[010]}$. (c),(d) Temperature dependence of (c) resistivity at 0~T for $I \parallel [100]$ and (d) magnetization at 0.1~T for $H \parallel [001]$ under various stresses. The data were taken during the warming process. In panel (c), each data except for 0~MPa is vertically shifted by 1~$\mu \Omega$ for clarity. (e),(f) Magnetic-field dependence of (e) magnetization and (f) its field derivative at 4~K for $H \parallel [001]$ under various stresses. The data were taken during the field-increasing process. Each dashed line in panel (e) traces the magnetization plateau associated with phases II, III, and II'. In panel (f), each data except for 0~MPa is vertically shifted for clarity.}
\label{Fig2}
\end{figure}

\begin{figure*}[t]
\centering
\includegraphics[width=0.96\linewidth]{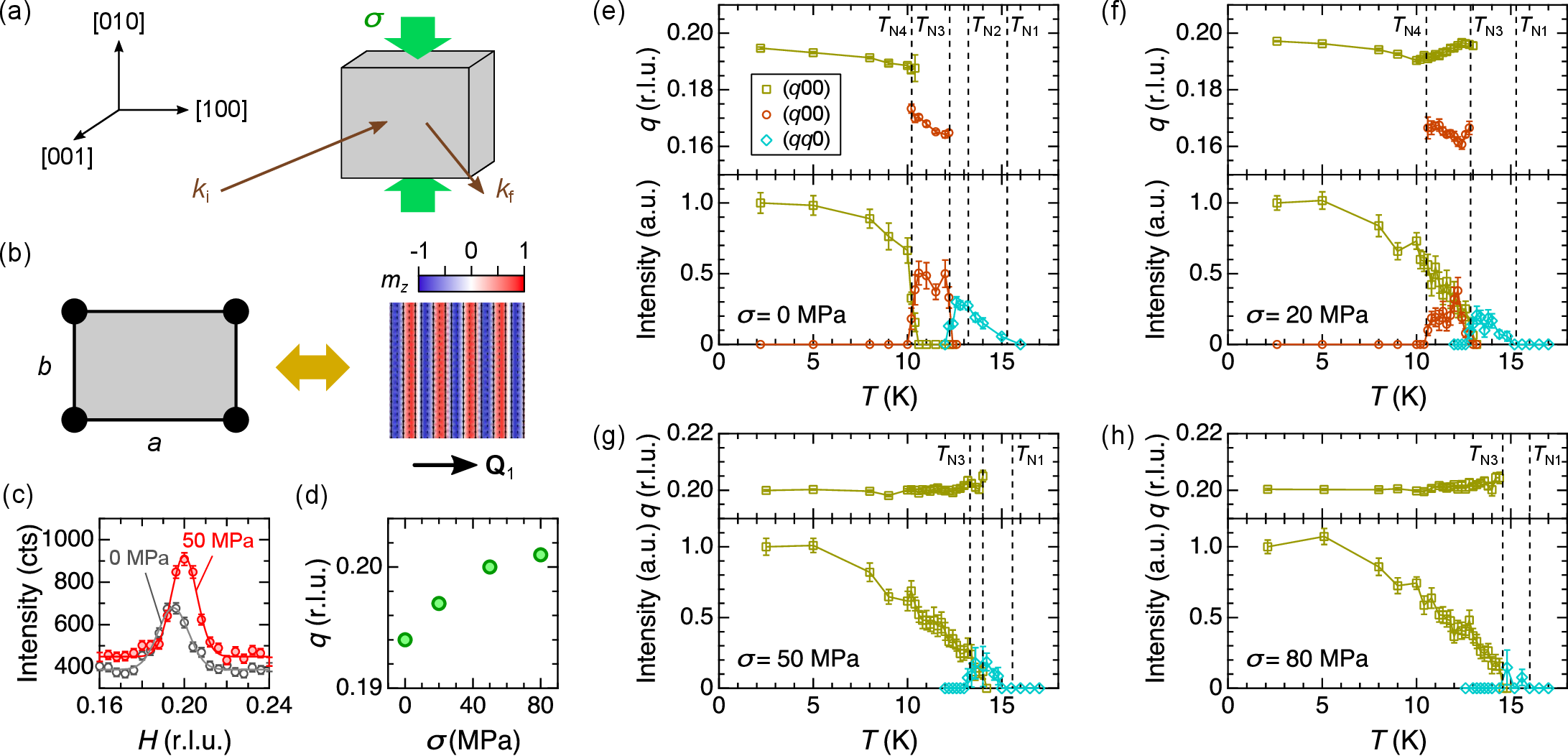}
\caption{(a) Schematic illustrations of the experimental geometry for neutron-scattering experiment under compressive stress $\sigma_{[010]}$. ${\mathbf k}_{i}$ (${\mathbf k}_{f}$) represents the propagation vector of the incident (scattered) neutron beam. (b) Correspondence between the orthorhombic distortion and the orientation of the ${\mathbf Q}$ vector in the spiral states of phase~I \cite{2023_Gen}. (c) Neutron-scattering profiles in the $(H, 0, 4)$ scan at $\sigma_{[010]}=0$ (gray) and $\sigma_{[010]}=50$~MPa (red). The data were collected at 2~K in zero magnetic field. (d) Stress dependence of the wavenumber $q$ at 2~K in phase~I. (e)--(h) Temperature dependence of $q$ (top) and the integrated intensities of magnetic Bragg peaks (bottom) in phases~I (yellow), V (orange), and VI/VII (cyan). The integrated intensities are estimated from Gaussian fits to the observed scattering profiles obtained in the $(H, 0, 4)$ or $(H, H, 4)$ scans, and then normalized to that of the $(q00)$ peak at 2~K.}
\label{Fig3}
\end{figure*}

Single crystals of EuAl$_{4}$, grown by the Al self-flux method \cite{2022_Tak}, were cut into rectangular parallelepiped shapes.
We used home-built probes equipped with a clamp-type uniaxial-stress cell \cite{2011_Nak, 2015_Nakajima, 2015_Nii, 2018_Nak, 2023_Sai} to apply compressive stress $\sigma_{[010]}$ along the [010] direction.
The resistivity was measured by a standard four-probe method with current $I \parallel [100]$ under vertical stress $\sigma_{[010]}$ [Fig.~\ref{Fig2}(a)], whereas the magnetization was measured under horizontal stress $\sigma_{[010]}$ with $H \parallel [001]$ [Fig.~\ref{Fig2}(b)].
Neutron scattering experiments were performed under vertical stress $\sigma_{[010]}$ with a triple-axis spectrometer (PONTA, 5G) at JRR-3, Japan Atomic Energy Agency \cite{2024_Nak}, with an incident neutron wavelength of 1.64~$\AA$ and the $(H, 0, L)$ scattering plane [Fig.~\ref{Fig3}(a)].
We observed magnetic Bragg peaks around the fundamental reflection at (0, 0, 4).
The out-of-plane peak at $(q, q, 4)$ ($q \approx 0.085$) was accessed by tilting the cryostat by $\sim$2.5$^{\circ}$, while the peak at $(0, q, 4)$ ($q \approx 0.19$) could not be accessed because of the limited tilt range.
In all measurements, the stress was applied or changed at 20~K in the PM phase.
Further experimental details are provided in the Supplemental Material \cite{SM}.

Figures~\ref{Fig2}(c) and \ref{Fig2}(d) show the temperature dependences of the resistivity $\rho(T)$ and magnetic susceptibility $\chi(T)$, respectively, measured under various stresses $\sigma_{[010]}$.
Under ambient pressure ($\sigma_{[010]} = 0$), $\rho(T)$ exhibits a kink at $T_{\rm N1} = 15.4$~K and a drop at $T_{\rm N3} = 12.2$~K, the latter coinciding with a tetragonal-to-orthorhombic structural transition \cite{2019_Shi, 2023_Gen}.
Clear anomalies appear in $\chi(T)$ also at $T_{\rm N2} = 13.2$~K and $T_{\rm N4} = 10.2$~K, whereas no corresponding features are resolved in $\rho(T)$ within our measurement sensitivity.
With increasing $\sigma_{[010]}$, anomalies in $\rho$ at $T_{\rm N1}$ and $T_{\rm N3}$ shift systematically to higher temperatures.
It is noteworthy that $\chi$ is suppressed under stress in phase~I, indicating an enhancement of antiferromagnetic character in the single-${\mathbf Q}$ spiral state.
This trend is consistent with the shortened magnetic modulation period revealed by neutron scattering, as discussed below.

Figure~\ref{Fig2}(e) shows the $M$-$H$ curves at 4~K under various $\sigma_{[010]}$ with $H \parallel [001]$.
At $\sigma_{[010]} = 0$, a sequence of magnetic transitions at $H_{\rm c1}$, $H_{\rm c2}$, and $H_{\rm c3}$ (from phases~I, II, and III to IV) is characterized by step-like increases in magnetization, which manifest as three pronounced peaks in $dM/dH$, marked by open triangles in Fig.~\ref{Fig2}(f).
Even at $\sigma_{[010]} = 10$~MPa, each $dM/dH$ peak shift slightly to higher fields and broadens.
Upon further increase of $\sigma_{[010]}$, the $dM/dH$ peak at $H_{\rm c1}$ splits into two, whereas the two peaks at $H_{\rm c2}$ and $H_{\rm c3}$ progressively coalesce and ultimately merge into a single peak at $\sigma_{[010]} = 160$~MPa.
The former behavior suggests the emergence of a new phase (phase~II') between phases~I and II, whereas the latter indicates the disappearance of phase~III [Fig.~\ref{Fig1}(a)].
This interpretation is further supported by the stress evolution of the plateaulike magnetization behavior in each phase, as indicated by the dashed lines in Fig.~\ref{Fig2}(e).
We also find that the upper critical field of phase~IV, $H_{\rm c4}$, increases monotonically with $\sigma_{[010]}$, whereas the lower critical field $H_{\rm c3}$ first rises up to $\sigma_{[010]} = 80$~MPa and then decreases at $\sigma_{[010]} = 160$~MPa.
This behavior suggests that an enhancement of the orthorhombic lattice distortion destabilizes the SkL and instead stabilizes the VL state, in agreement with a recent theoretical prediction by Hayami~\cite{2025_Hay}.

To validate the accuracy of our uniaxial-stress measurements and the phase diagrams derived from them (Fig.~\ref{Fig1}), we here compare the observed stress dependence of the phase boundaries with a thermodynamic estimate.
As the magnetic transition at $T_{\rm N3}$ is first order, its stress dependence is governed by a Clausius-Clapeyron relation, $dT_{\rm N3}/d\sigma_{[010]} = V (\Delta L/L_{[010]})/{\Delta S}$, where $V$ is the volume per formula unit, $\Delta L/L_{[010]}$ is the linear thermal expansion along [010], and $\Delta S$ is the entropy change at $T_{\rm N3}$.
Using literature values $V = 107.1$~\AA$^{3}$/f.u. (at 20~K) \cite{2022_Ram}, $\Delta L/L_{[010]} \approx 4 \times 10^{-4}$ \cite{2022_Mei, 2023_Gen}, and $\Delta S \approx 1$~J/K$\cdot$mol \cite{2015_Nak}, we obtain $dT_{\rm N3}/d\sigma_{[010]} \approx 30$~K/GPa.
This estimate is consistent with the experimental slope between 0 and 80~MPa, $dT_{\rm N3}/d\sigma_{[010]} \approx 25$~K/GPa [Fig.~\ref{Fig2}(c)].
Similarly, the stress dependence of the upper critical field of phase~I follows $d(\mu_{0}H_{\rm c1})/d\sigma_{[010]} = V (\Delta L/L_{[010]})/{\Delta M}$, where $\Delta L/L_{[010]}$ is the linear magnetostriction along [010] and $\Delta M$ is the magnetization jump at $H_{\rm c1}$.
Using $\Delta L/L_{[010]} \approx 5 \times 10^{-4}$ \cite{2022_Mei, 2023_Gen} and $\Delta M \approx 2$~$\mu_{\rm B}$/f.u. [Fig.~\ref{Fig2}(e)], we obtain $d(\mu_{0}H_{\rm c1}/d\sigma_{[010]}) \approx 3$~T/GPa, also in good agreement with experiment.
It is worth noting that the stress dependence of the critical temperatures and fields of the magnetic phases in EuAl$_{4}$ exceeds, by about two orders of magnitude, thermodynamic estimates for another SkL host Gd$_{2}$PdSi$_{3}$ \cite{2021_Spa}.
The pronounced stress sensitivity of the magnetic phase diagram in EuAl$_{4}$ is consistent with its unusually large thermal expansion and magnetostriction compared with conventional rare-earth materials \cite{2019_Shi, 2022_Mei, 2023_Gen}.

We next turn to the evolution of the magnetic structures under stress in zero field.
A previous resonant x-ray scattering revealed that the ${\mathbf Q}$ vector runs along the elongated $a$ axis in phase~I \cite{2023_Gen} [Fig.~\ref{Fig3}(b)].
Accordingly, under vertical compressive stress $\sigma_{[010]}$, the orthorhombic domains are expected to align such that the magnetic Bragg reflections lie within the horizontal scattering plane.
This scenario is confirmed by the comparison of neutron-scattering profiles in the $(H, 0, 4)$ scan at 2~K without and with applied stress, as shown in Fig.~\ref{Fig3}(c); the peak intensity nearly doubles in the stressed case.
In addition, a shift of the peak position is clearly observed.
Figure~\ref{Fig3}(d) shows the stress dependence of the magnetic modulation wavenumber $q$ at 2~K: $q = 0.194$ at $\sigma_{[010]} = 0$, consistent with previous reports \cite{2021_Kan, 2022_Tak, 2023_Gen, 2024_Mia, 2024_Vib}, while it increases to 0.197 at 20~MPa and 0.200 at 50~MPa.
At $\sigma_{[010]} = 80$~MPa, $q$ further increases to 0.201, but the change becomes smaller, suggesting a tendency toward saturation.

Figures~\ref{Fig3}(e)--\ref{Fig3}(h) summarize the temperature dependence of $q$ for the $(q00)$ peak(s) and the integrated intensities of the $(q00)$ and $(qq0)$ peaks under various $\sigma_{[010]}$.
We note that no stress effect on the $q$ value of the $(qq0)$ peak was detected in phases VI and VII within the experimental resolution.
At $\sigma_{[010]} = 0$, $q$ decreases gradually from 0.194 to 0.187 in phase~I with increasing temperature up to $T_{\rm N4}$, followed by a discontinuous jump to $q \approx 0.17$, associated with the phase transition to phase~V.
The ${\mathbf Q}$ vector then switches from $(q, 0, 0)$ ($q \approx 0.165$) to $(q, q, 0)$ ($q \approx 0.085$) at $T_{\rm N3}$.
In the present experiments, we could not observe the high-harmonic modulation corresponding to ${\mathbf Q}_{1} + {\mathbf Q}_{2}$ because of insufficient intensity resolution, and therefore could not distinguish between the double-${\mathbf Q}$ ML and VL states identified in phases VI and VII, respectively \cite{2022_Tak}.
Remarkably, at $\sigma_{[010]} = 20$~MPa, phase~V coexists with phase~I in the temperature range between  $T_{\rm N4} = 10.6$~K and $T_{\rm N3} = 12.8$~K.
At $\sigma_{[010]} = 50$~MPa, phase V eventually disappears, and phase I remains stable up to $T_{\rm N3} = 13.4$~K.
Given that the orthorhombic lattice distortion is larger in phase~I than in phase~V \cite{2019_Shi, 2023_Gen}, the disappearance of phase~V may be attributed to the rapid increase in $T_{\rm N4}$ with $\sigma_{[010]}$, i.e., a large $dT_{\rm N4}/d\sigma_{[010]}$, which is likely due to the tiny entropy change at the phase boundary \cite{2015_Nak}.
The spiral states in phases I and V have been proposed to possess opposite chiralities \cite{2024_Mia, 2024_Vib}.
Our present results suggest that the chirality of the spiral order can be controlled by uniaxial stress in the intermediate temperature range of 10--12 K.

\begin{figure}[t]
\centering
\includegraphics[width=\linewidth]{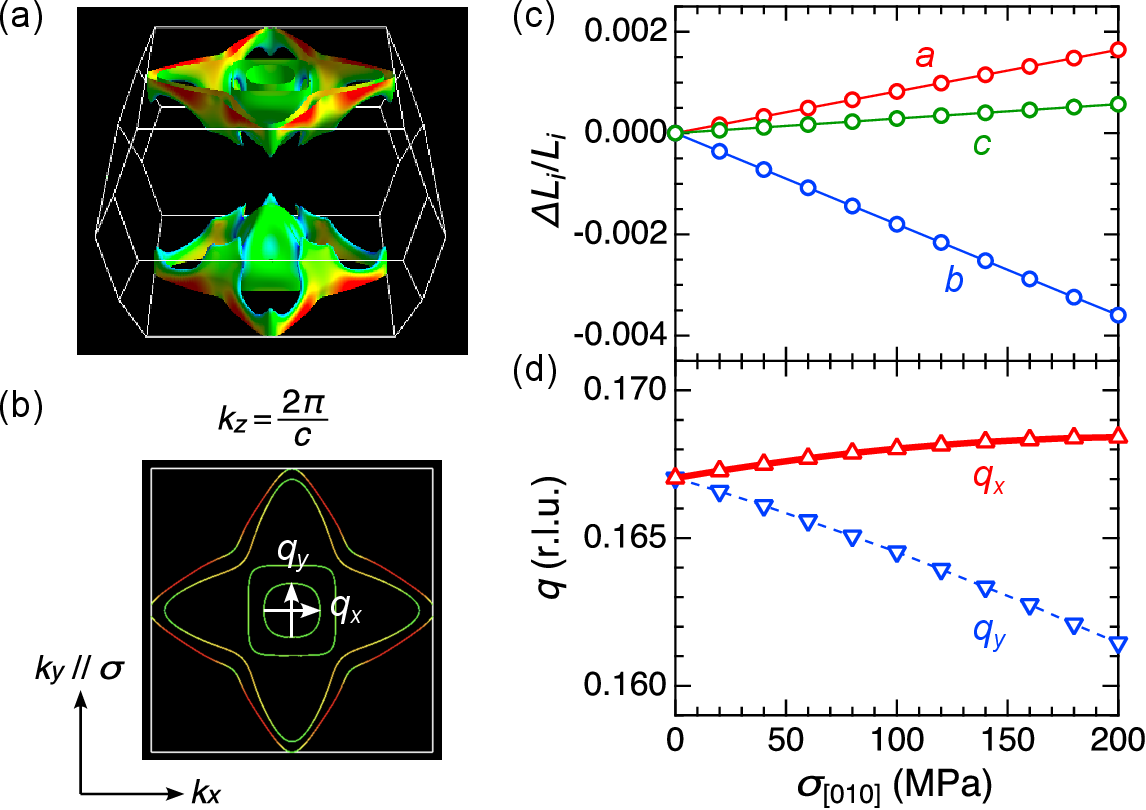}
\caption{(a),(b) Calculated Fermi surfaces (FSs) for band \#1 of EuAl$_{4}$, which is relevant to the magnetism. Panel (b) shows the FSs near the Z point within the $k_{x}$-$k_{y}$ plane. Illustrations are drawn by the \textsc{fermisurfer} code \cite{2019_Kaw}. (c),(d) Calculated $\sigma_{[010]}$ dependence of (c) the lattice constants and (d) the nesting vectors $q_{x}$ and $q_{y}$, as defined in panel (b).}
\label{Fig4}
\end{figure}

To understand the evolution of $q$ with $\sigma_{[010]}$ in phase~I [Fig.~\ref{Fig3}(d)], we investigate how the Fermi surfaces (FSs) are modified by $\sigma_{[010]}$ using first-principles calculations.
We take the tetragonal structure (space group  $I4/mmm$) as the starting point \cite{comment}.
We find two bands, \#1 and \#2, crossing the Fermi level, both exhibiting three-dimensional character.
Band \#1 produces two star-shaped and two square-shaped FSs around the Z point [Figs.~\ref{Fig4}(a) and \ref{Fig4}(b)], while band \#2 yields two star-shaped FSs around the $\Gamma$ point (see the Supplemental
Material \cite{SM}), in agreement with previous angle-resolved photoemission spectroscopy (ARPES) and first-principles calculations \cite{2024_Mia, 2016_Kob, 2024_Eat}.
The calculated nesting vector near Z is $q_{x} = q_{y} = 0.167$ at $\sigma_{[010]} = 0$ [Fig.~\ref{Fig4}(b)], which roughly matches the ${\mathbf Q}$ vector in phase~V but slightly smaller than that in phase~I.
Since the calculated nesting vector near $\Gamma$ largely deviate from the experimental ${\mathbf Q}$ vector \cite{SM}, the FS nesting around Z is likely more relevant to the RKKY interaction in EuAl$_{4}$.
Our calculations show that the application of compressive stress $\sigma_{[010]}$ induces the change in the lattice constants as shown in Fig.~\ref{Fig4}(c).
Concomitantly, the nesting vector $q_{x}$ (perpendicular to $\sigma_{[010]}$) increase by 0.001, whereas $q_{y}$ decreases by 0.025 at 100~MPa [Fig.~\ref{Fig4}(d)].
The increase in $q_{x}$ is qualitatively consistent with the magnetic structure change in phase~I, given that the helical modulation runs perpendicular to the compressed [010] axis.
However, the calculated rate of increase in $q_{x}$ is about an order of magnitude smaller than the observed increase of $q$.
This discrepancy likely stems from our neglect of the orthorhombic lattice distortion that develops below $T_{\rm N3}$ and is further enhanced below $T_{\rm N4}$ \cite{2019_Shi, 2023_Gen}.
Even at $\sigma_{[010]} = 0$, the orthorhombic distortion reaches $\sim$2\% below 5~K \cite{2023_Gen}, comparable to the strain induced by $\sim$100~MPa in our calculations [Fig.~\ref{Fig4}(c)].
Moreover, a recent ARPES study revealed FS reconstruction in phase~I \cite{2024_Eat}, suggesting that the orthorhombic distortion can modify the nesting conditions.
Incorporating the spontaneous lattice distortion is therefore expected to amplify the stress-induced anisotropy of the FS and reconcile the calculations with experiment.

The CDW in EuAl$_{4}$ has been proposed to originate primarily from electron-phonon coupling rather than the FS nesting \cite{2025_Cao, 2025_Suk}.
Synchrotron x-ray diffraction studies have hinted at a slight orthorhombic distortion already below $T_{\rm CDW}$ \cite{2022_Ram, 2024_Kor, 2025_Kot}.
In our first-principles calculations, imposing a commensurate modulation with ${\mathbf Q}_{\rm CDW} = (0, 0, 1/6)$ does stabilize the CDW state, but we find no intrinsic tendency for an orthorhombic distortion to be stabilized.
This suggests that, in addition to the CDW, spin-lattice coupling is likely essential for driving the tetragonal-to-orthorhombic transition at $T_{\rm N3}$.
The intertwined spin, lattice, and charge degrees of freedom would render the FS highly sensitive to uniaxial stress, thereby enabling decisive control of the phase diagram and magnetic structures.

In summary, we have demonstrated that compressive uniaxial stress $\sigma_{[010]}$ substantially alters the stability of the versatile helimagnetic phases in EuAl$_{4}$, including the rhombic and square SkL phases (Fig.~\ref{Fig1}), using resistivity and magnetization measurements complemented by neutron scattering.
The increases of the critical temperatures and fields with $\sigma_{[010]}$ are attributable to an enhanced antiferromagnetic character, evidenced by the suppression of magnetic susceptibility and the shortened magnetic modulation period (i.e., increased $q$) in the lowest-temperature single-${\mathbf Q}$ spiral state.
The growth of $q$ with $\sigma_{[010]}$ can be traced to Fermi-surface deformation induced by orthorhombic strain, as supported by our first-principles calculations.
The present mechanism for uniaxial-stress control of helimagnetism, rooted in the FS instability toward orthorhombic distortion, contrasts with the conventional approach in chiral magnets that relies mainly on tuning magnetic anisotropy and/or the DM interaction \cite{2015_Nii, 2015_Cha, 2017_Fob, 2015_Shi, 2020_Bud, 2022_Lit, 2017_Sek, 2018_Nak}, thereby paving the way to manipulate physical properties in spin--lattice--charge-coupled systems.

The neutron-scattering experiment
at JRR-3 was carried out by Proposals No.~23801 and No.~24508.
This work was financially supported by the JSPS KAKENHI Grants-In-Aid for Scientific Research (Grants No.~21H04990, No.~22H04965, No.~23K13068, No.~24H02235, and No.~25H00611), JST CREST (Grant No.~JPMJCR23O4), Murata Science Foundation, and Asahi Glass Foundation.

\end{document}